\documentstyle[12pt,prd,aps]{revtex}
\def\be{\begin{equation}}
\def\en{\end{equation}}
\def\bea{\begin{eqnarray}}
\def\ena{\end{eqnarray}}
\def\gsim{\mathrel{\raise.3ex\hbox{$>$}\mkern-14mu
             \lower0.6ex\hbox{$\sim$}}}
\def\lsim{\mathrel{\raise.3ex\hbox{$<$}\mkern-14mu
             \lower0.6ex\hbox{$\sim$}}}

\begin{document}
\title{Topological defects and open inflation} 
\author{Alexander Vilenkin} \address{Institute of Cosmology, Department of
Physics and Astronomy,\protect\\ Tufts University, Medford, MA
02155, USA} \maketitle

\begin{abstract}
Topological defects can naturally be formed soon after bubble
nucleation in the open inflation scenario.  The defects are not
completely diluted away by the subsequent period of inflation in the
bubble interior and can produce observable large-scale microwave
background anisotropies.  Superheavy strings and monopoles attached to
the strings can act as gravitational lenses with angular separation
between the images of up to an arc minute.
\end{abstract}
\pacs{1}

\section{Introduction}

Until recently, a flat universe with $\Omega=1$ was regarded as a firm
prediction of inflationary models.  However, the observational
evidence for a flat universe is far from being clear, and a new class
of models called ``open inflation'' has recently
become a subject of active investigation \cite{Gott,Bucher,LM}.  
In these models,
the visible universe is contained within a single bubble which
nucleated in an inflating false vacuum.  The observed homogeneity of
the universe is ensured by a high symmetry of the bubble, rather
than by a large expansion factor, as in the more familiar inflation
models \cite{Linde}.  After nucleation, inflation continues for some time
in the bubble interior, and the corresponding number of e-foldings can
be adjusted to give any value of the density parameter $\Omega$
between $\Omega=0$ and $\Omega=1$.  

The first models of open inflation \cite{Gott,Bucher} gave a fixed value of
$\Omega$, and obtaining a value not too close to 0 or 1
required a substantial amount of fine-tuning.  A significant
improvement was made by Linde and Mezhlumian \cite{LM} who suggested a
number of models where  $\Omega$ is a continuous parameter taking
different values in different bubbles.  The probability distribution
for $\Omega$ in such models was discussed in Ref. \cite{VW}, where it
was argued that, with the anthropic factor properly taken into
account, this distribution can naturally be peaked at an intermediate
value of $\Omega$.  Open inflation may thus require no more
fine-tuning than `ordinary' inflation.

The purpose of this paper is to make a simple observation that, 
in models of open inflation, topological
defects are likely to be formed soon after the bubble nucleation.
The defects will not be completely diluted away by the short period of
inflation in the bubble interior, and we may still be able to observe
them on very large scales comparable to the curvature scale of the
universe. 

\section{Defect formation and evolution}

The models suggested by Linde and Mezhlumian involve two scalar
fields, $\phi_1$ responsible for bubble nucleation and $\phi_2$
responsible for the slow-roll inflation inside the bubble.  For a
specific example consider the model with a potential \cite{LM}
\begin{equation}
U(\phi_1,\phi_2)=V(\phi_1) + \lambda\phi_1^2 \phi_2^2,
\label{1}
\end{equation}
where $V(\phi_1)$ has a metastable minimum at $\phi_1=0$ and the true
minimum at $\phi_1=\eta_1$.  The stage for open inflation is set by
the inflating false vacuum at $\phi_1 =0$.  The corresponding energy
density is $\rho_1=V(0)$, and the spacetime is nearly de Sitter with
an expansion rate $H_1=(8\pi\rho_1/3m_p^2)^{1/2}$, where $m_p$ is the
Planck mass.  Bubble nucleation can occur to a wide range of values of
the field $\phi_2$.  The second stage of inflation takes place while
$\phi_2$ rolls towards its minimum at $\phi_2=0$, and the resulting
density parameter $\Omega$ is determined by the initial value of
$\phi_2$ at the time of bubble nucleation.  

The metric in the bubble interior is well approximated by the
Robertson-Walker (RW) metric
\begin{equation}
ds^2=dt^2-a^2(t)[d\zeta^2+\sinh^2\zeta (d\theta^2+\sin^2\theta d\phi^2)]
\label{2}
\end{equation}
with $a(t)$ satisfying
\begin{equation}
{\dot a}^2-1=(8\pi/3m_p^2)\rho a^2.
\label{3}
\end{equation}
Here, $\rho$ is the energy density measured by co-moving RW observers.
The hypersurface $t=0$, where $a=0$, is the future light cone of the
bubble center at the moment of nucleation.  The coordinate system
(\ref{2}) has a singularity on this surface, but the four-geometry is
of course non-singular.

For a generic shape of the potential $V(\phi_1)$, the initial radius
of the bubble $R$ is comparable to the thickness of the bubble wall.
At $t\sim R$, when $a(R)\sim R$ and $\rho\sim\rho_1$, the field
$\phi_1$ begins to oscillate about $\phi_1=\eta_1$, and at later times
its energy density scales like that of non-relativistic matter.  The
initial bubble radius cannot exceed the horizon, $R\lsim H_1^{-1}$,
and it is easily seen from Eq. (\ref{3}) that the density term is
either negligible from the very beginning or becomes
negligible in a few Hubble times after the nucleation.  Hence, the
buble-universe is curvature-dominated, with $a(t)\approx t$ and
$\rho\sim\rho_1(R/t)^3$.  

The second period of inflation begins at $t\sim H_2^{-1}$, where
$H_2=(8\pi\rho_2/3m_p^2)^{1/2}$, $\rho_2=\lambda\eta_1^2\phi_{2i}^2$
and $\phi_{2i}$ is the initial value of $\phi_2$ at nucleation.
Between the nucleation and second inflation, the bubble-universe
expands by a factor $f\sim(H_2R)^{-1}\gsim
H_1/H_2 \sim (\rho_1/\rho_2)^{1/2}$.  The value of $\rho_2$ is
constrained to be smaller than $\rho_1$, and since there is no reason
for $\rho_2$ to be $\sim\rho_1$, we expect that 
generically $\rho_2 \ll \rho_1$ and $f\gg 1$.  The field
$\phi_2$ evolves on a timescale $H_2/m_2^2\gg H_2^{-1}$ and remains
essentially constant during the period between the two inflations.
Here, $m_2=(2\lambda)^{1/2}\eta_1\ll H_2$ is the mass of $\phi_2$.

Defect formation between inflations can be triggered by the same  
mechanisms as defect formation during inflation which has been
previously discussed in the literature \cite{Shafi,KL,VOS,Book}.  
Let $\chi$ be the field responsible for the defects and
let us first consider a model where $\chi$ has a non-minimal
coupling to the curvature ${\cal R}$,
\begin{equation}
L_\chi=|\partial_\mu\chi|^2+M^2|\chi|^2 -\kappa {\cal R}|\chi|^2
-\lambda_\chi |\chi|^4.  
\label{4}
\end{equation}
We shall assume for definiteness that $\chi$ is a complex scalar
field.  The effective mass squared of $\chi$ is
\begin{equation}
m_\chi^2=-M^2 +\kappa{\cal R}.
\label{5}
\end{equation}
From Einstein's equations ${\cal R}= 8\pi\rho/m_p^2$, and thus  $m_\chi^2$
changes sign at $\rho_c=M^2 m_p^2/8\pi\kappa$.  This triggers a
symmetry-breaking phase transition resulting in the formation of
cosmic strings with mass per unit length $\mu\sim M^2/\lambda_\chi$.
The phase transition occurs between inflations provided that
$\rho_1>\rho_c>\rho_2$ and $M>H_2$.  These conditions require no fine
tuning and can be naturally satisfied in a variety of models.  

An alternative scenario is that the $\phi_2$ particles are unstable
with a lifetime $\tau_2< H_2^{-1}$.  Their decay products (call them
$\sigma$) may then trigger the phase transition by a direct interaction
with $\chi$, e.g., $L_{\chi\sigma}=-\lambda_{\chi\sigma}\sigma^2
|\chi|^2$.  If the decay products of $\phi_2$ have enough time to
thermalize, then the corresponding phase transition is the `usual',
thermal phase transition.

Finally, there is one more mechanism which is specific to open
inflation.  Suppose now that $\chi$ is coupled to the tunneling field
$\phi_1$,
\begin{equation}
L_\chi=|\partial_\mu\chi|^2 -M^2|\chi|^2+\lambda_1\phi_1^2|\chi|^2
-\lambda_\chi|\chi|^4. 
\label{6}
\end{equation}
In the false vacuum, $\phi_1=0$ and $\chi=0$, while in the true vacuum
$\chi$ has a non-zero expectation value.  Now, there are two
possibilities.  If a bubble nucleates with a
non-zero value of $\chi$, then this value will tend to be homogeneous
throughout the bubble-universe, and no defects will be formed.
Alternatively, $\chi=0$ at nucleation and defect formation does occur.
The initial value of $\chi$ is determined by the instanton solution of
the Euclidean field equations \cite{Coleman}.  
The four-geometry of this solution is
compact, with a characteristic size $\sim H_1^{-1}$.  If this size is
smaller than the length scales characterizing the field $\chi$,
$M^{-1}$ and $\lambda_1^{-1/2}\eta_1^{-1}$, then $\chi=0$ on the
instanton.  For $H_1>M>H_2$,  the field $\chi$ initially remains near
zero, and starts evolving at $t\sim M^{-1}$.  This can be regarded as
the time of string formation.

If strings are formed at $t_i\sim M^{-1}$, then the initial string
separation is $\xi_i\sim M^{-1}$, which corresponds to a co-moving
scale $\zeta\sim 1$ in the metric (\ref{2}).  
In the course of the following evolution,
$\xi$ is bounded from above by the horizon, $t$, and from below by
$[a(t)/a(t_i)]\xi_i$.  Since both of these bounds are $\sim t$, we have
$\xi\sim t$ all the way until the beginning of the second inflation,
$t\lsim H_2^{-1}$.  To be more precise, the horizon
in metric (\ref{2}) with $a(t)\propto t$ grows like $t\ln t$ and
Spergel \cite{Spergel} has argued that the characteristic scale of
defects will grow like 
\begin{equation}
\xi\propto t(\ln t)^{1/2}.
\label{6'}
\end{equation}  
In most of the following discussion I disregard these logarithmic factors. 

During inflation, the strings are ``frozen'', that is, they are
conformally stretched with their co-moving scale remaining at
$\zeta\sim 1$.  They start moving again at $t_*\sim\Omega/H_0$ (or
redshift $(1+z_*)\sim \Omega^{-1})$, when
the curvature scale $\zeta\sim 1$ comes within the horizon.  Here,
$H_0$ is the value of the Hubble parameter, $H={\dot a}/a$, at the
present time.  At $t>t_*$ the strings will evolve in a scaling
regime with $\xi(t)\sim t$, and thus the co-moving scale of strings
will remain comparable to the curvature scale.  
The growth of density fluctuations in an open universe ceases at $t>t_*$,
and since the strings do not generate fluctuations at $t<t_*$,
they clearly could not be responsible 
for structure formation.  Yet, the mass scale of the strings can be
quite high and they can produce some observable effects.

In models where the phase transition is triggered by the curvature or by
the decay products of $\phi_1$, it is possible to have $\xi_i\ll t_i$.
The strings will then evolve towards a
scaling regime with $\xi(t)\sim t$, and for $t_i\ll H_2^{-1}$ we
may still have $\xi\sim t$ by the beginning of second inflation.
However, the timescale on which the string scale grows up to $t$ is
very model-dependent.  For example, strings may be overdamped due to
their interaction with the decay products of $\phi_1$.  In this case
their evolution will be rather slow, and the co-moving scale of
strings during inflation may be $\zeta\ll 1$.  It is conceivable that
this scale can be small enough for strings to play a role in structure
formation.  

Suppose now that the defects formed between inflations are magnetic
monopoles.  The typical co-moving separation of the monopoles is then
$\zeta\sim 1$, and the resulting monopole density is totally
negligible.  However, the situation is drastically changed if the
monopoles get connected by strings at a subsequent phase transition.

Strings can either be formed during inflation or in the
postinflationary epoch.  The characteristic length scale of string at
formation is then much smaller than the monopole separation; the
strings connecting monopoles have Brownian shapes, and there is a
large number of closed loops.  The evolution of strings after
inflation is initially identical to that of topologically stable
strings, without monopoles \cite{Book}.  In the course of the evolution,
the characteristic length of strings grows like $t$ and becomes
comparable to the monopole separation at $t\sim t_*$, so that we are
left with monopole-antimonopole pairs connected by more or less
straight strings of length $\ell \sim t_*$.  At $t>t_*$, the pairs
oscillate and gradually lose their energy by gravitational or gauge
boson radiation.  The corresponding lifetimes are, respectively
\cite{MV,BMV}, $\tau_{gr}\sim t_*/\Gamma_{gr}G\mu$ and $\tau_b\sim
\mu t_*/\alpha m^2$.  Here, $\Gamma_{gr}\sim 10^3$ 
is a numerical coefficient, $\alpha\sim 10^{-2}$ is the gauge coupling
constant, and $m$ is the monopole mass.  Since the symmetry breaking
scale of the monopoles is greater than that of the strings, we should
have $m^2 \gsim\mu$, and it is easily seen that both
lifetimes $\tau_{gr}$ and $\tau_b$ are much greater than the present
age of the universe (provided that $\Omega
\gsim 0.1$).  Note that if monopoles and strings are
formed after inflation, then the lifetime of the pairs is typically
very short, and they decay well before the end of the radiation era. 

\section{Observational effects}

Let us first consider possible effects of `plain' cosmic strings,
without monopoles.
Strings can produce double images of high-redshift galaxies and
quasars \cite{V81}.  The metric of a straight string in an open universe
is given \cite{AFV} by the 
same Eq.(\ref{2}), but with a modified range for the angular variable
$\phi$, $0<\phi<2\pi-\Delta$, where $\Delta=8\pi G\mu$ is the deficit
angle and $\mu$ is the mass per unit length of string.  Let $\zeta_O$
and $\zeta_G$ be, respectively, the coordinate distances from the
string to the observer and to the galaxy, and let $\theta$ be the angle
between the string and the line of sight.  Then it is easily shown
that the angular separation between the images is 
\begin{equation}
\delta=8\pi G\mu{\sinh \zeta_G \over{\sinh
(\zeta_G+\zeta_O)}}\sin\theta.
\label{7}
\end{equation}
Here, I have assumed for simplicity that the string is static and that
$G\mu\ll 1$.  For $\zeta_G,\zeta_O\ll 1$, Eq. (\ref{7}) reduces to the
expression for the angular separation in a spatially flat universe
\cite{V81}. 

Moving strings induce discontinuous jumps in the microwave background
temperature \cite{KS,G85},
\begin{equation}
\delta T/T\sim 8\pi\beta G\mu v,
\label{8}
\end{equation}
where $v$ is the string velocity with respect to local co-moving
observers and $\beta\sim 0.5$ is a trigonometric factor.  
At $t>t_*$, the universe becomes curvature-dominated and
expands like $a(t)\propto t$.  This regime is ``on the verge'' of
inflation, in the sense that an expansion law $a(t)\propto t^\alpha$
with $\alpha>1$ corresponds to a power-law inflation.  Since the
strings are frozen during inflation, one can expect \cite{PS} their
typical velocity in a curvature-dominated universe to be well below
that in a radiation or matter-dominated universe ($v_r\sim v_m\sim
0.6$).   

The pattern of the microwave temperature fluctuations on the sky is a
superposition of the contributions of strings from different redshifts
between $z_*\sim\Omega^{-1}$ and the present.  The highest density of
strings corresponds to the largest redshifts near $z_*$, and the typical
angular separation of discontinuities on the sky is
$\theta_{min}\sim\Omega$.  We expect temperature fluctuations of
magnitude (\ref{8}) on scales $\theta_{min}\lsim
\theta {\buildrel <\over\sim}1$.  To avoid conflict with observations,
we should require $\delta T/T {\buildrel <\over\sim}10^{-5}$.  With
$v\sim 0.2$ this gives a bound $G\mu \lsim 10^{-5}$.
According to Eq. (\ref{7}), a string with $G\mu\sim 10^{-5}$ can give
double images with angular separations of up to an arc minute.

Gravitational waves produced by the strings in this scenario will have
too low frequencies to be detected by the existing methods, and thus 
the usual bounds on $G\mu$ from
the millisecond pulsar observations and from nucleosynthesis
considerations do not apply.

Note that if the logarithmic factor in Eq. (\ref{6'}) is indeed
present, then the co-moving scale of strings at $t<t_*$ can be
somewhat larger than the curvature scale.  This will have the effect
of moving $t_*$ closer to the present time and increasing the angle
$\theta_{min}$.  In extreme cases, the typical string separation can
even be greater than the present horizon.

Let us now turn to monopoles connected by strings.  Monopoles are
pulled by the strings with a force $\mu$ and accelerate to
ultrarelativistic speeds.  The typical energy of a monopole at time
$t<t_*$ is $E\sim\mu t$, and at $t>t_*$ is $E\sim\mu t_*$.  The latter
corresponds to a mass
\begin{equation}
\mu t_*\sim 10^{16}\left({G\mu\over{10^{-6}}}\right)\Omega M_\odot.
\label{9}
\end{equation}
For grand-unification-scale strings with $G\mu\sim 10^{-6}$, this is
comparable to the mass of a supercluster!

The microwave anisotropy produced by a moving point mass in Minkowski
space has been calculated by Stebbins \cite{Stebbins},
\begin{equation}
{\delta T\over{T}}=-{4GEv_\bot\over{r\theta}}\cos\alpha.
\label{10}
\end{equation}
Here, $r$ is the distance from the observer to the point mass,
$E=M(1-v^2)^{-1/2}$ is the total energy of the mass, $v$ is its velocity,
$v_\bot$ is the velocity projected on the sky, $\theta$ is the angular
distance between the mass ($M$) and the point of
observation ($P$), and $\alpha$ is the angle between the line $MP$ and
the projected velocity.  It is assumed that $\theta\ll 1$.  
To extend this equation to the case of an expanding universe, we note
that $r\theta= \ell_\bot$ is the closest-approach distance between the
microwave photon and the mass.  In a low-density universe, and for a
mass at a redshift in the range $1\lsim z\lsim  
z_*$, this relation is replaced by \cite{Refp2}
$\theta=H_0 \ell_\bot F(z)$, where $F(z)\approx 2$.  Hence, we can write
\begin{equation}
\delta T/T=8GH_0Ev_\bot \theta^{-1}\cos\alpha.
\label{11}
\end{equation}
With $E\sim\mu t_*$ and $v_\bot\sim 1$, this gives 
\begin{equation}
\delta T/T\sim 4G\mu\Omega \theta^{-1}.
\label{12}
\end{equation}
The density of monopoles with $z\lsim z_*$ on the sky is
dominated by the monopoles at $z\sim z_*$, so the typical angular
separation between monopoles is $\theta\sim\Omega$.  Monopoles at
$z>z_*$ have smaller energies, and the corresponding microwave
anisotropies are proportionally weaker.  For $G\mu\sim 10^{-6}$ and
$\Omega \gsim 0.1$, the anisotropies produced by
moving monopoles have a detectable magnitude $\delta T/T\gsim  
10^{-5}$ up to the angular distance of a few degrees.

Ultrarelativistic monopoles can also act as gravitational lenses.  If
the monopole and the galaxy which is being lensed are both at $z\sim
1$, then the typical light deflection angle is $\theta\sim
GE/\ell_\bot$.  Since $\ell_\bot\sim\theta H_0^{-1}$ and $E\sim\mu
t_*$, we obtain an estimate for the typical angular separation between
the images,
\begin{equation}
\delta\phi\sim (G\mu\Omega)^{1/2}.
\label{13}
\end{equation}
For $G\mu\sim 10^{-6}$ this separation is about an arc minute. 

Global monopoles and textures produced between inflations will
generate microwave background anisotropies, and global monopoles can
also act as gravitational lenses.
The evolution of global defects in an open universe and their effect
on the microwave background have
been studied by Pen and Spergel \cite{PS}.  A bound on the symmetry
breaking scale of defects can probably be extracted from their
results, but I was unable to do so.

Finally, I would like to emphasize that, apart from defect formation
between inflations, some topological defects can be formed during or
after inflation.  The characteristic length scale of such defects is
much smaller than the curvature scale, and they may be suitable as
seeds for structure formation.  Defect-seeded structure formation in
an open universe has been discussed in Refs. \cite{PS,Ferreira}.

I am grateful to David Spergel and Albert Stebbins for
useful correspondence and to Martin Bucher and Andrei Linde for their
comments on the paper.  This work
was supported in part by the National Science Foundation.


\begin{thebibliography}{99}

\bibitem{Gott} J.R. Gott, Nature {\bf 295}, 304 (1982).

\bibitem{Bucher} M. Bucher, A.S. Goldhaber and N. Turok,
Phys. Rev. {\bf D55}, 3314 (1995);  K. Yamamoto, M. Sasaki and
T. Tanaka, Ap. J. {\bf 455}, 412 (1995).

\bibitem{LM} A.D. Linde, Phys. Lett. {\bf B351}, 99 (1995);
A.D. Linde and A. Mezhlumian, Phys. Rev. {\bf D52}, 6789 (1995).

\bibitem{Linde} For a review of inflation see A.D. Linde, {\it
Particle Physics and Inflationary Cosmology} (Harwood, Chur,
Switzerland, 1990); K.A. Olive, Phys. Rep. {\bf 190}, 307 (1990).

\bibitem{VW} A. Vilenkin and S. Winitzki, Phys. Rev. {\bf D55}, 548
(1997).

\bibitem{Shafi} Q. Shafi and A. Vilenkin, Phys. Rev. {\bf D29}, 1870
(1984).

\bibitem{KL} L. Kofman and A.D. Linde, Nucl. Phys. {\bf B282}, 555
(1987).

\bibitem{VOS} E.T. Vishniak, K.A. Olive and D. Seckel,
Nucl. Phys. {\bf B289}, 717 (1987).

\bibitem{Book} For a review of topological defects in cosmology, see
M.B. Hindmarsh and T.W.B. Kibble, Rep. Prog. Phys. {\bf 55}, 478
(1995); A. Vilenkin and E.P.S. Shellard, {\it Cosmic Strings and Other
Topological Defects} (Cambridge University Press, Cambridge, 1994).

\bibitem{Coleman} S. Coleman and F. de Luccia, Phys. Rev. {\bf D21},
3305 (1980).

\bibitem{Spergel} D.N. Spergel, Ap. J. Lett. {\bf 412}, 5 (1993).

\bibitem{MV} X. Martin and A. Vilenkin, Phys. Rev. Lett. {\bf 77},
2879 (1996); Phys. Rev. D, in press [gr-qc/9612008].

\bibitem{BMV} V. Berezinsky, X. Martin and A. Vilenkin, unpublished
[astro-ph/9703077].

\bibitem{V81} A. Vilenkin, Phys. Rev. {D23}, 852 (1981);
Ap. J. Lett. {\bf 51}, 282 (1984).

\bibitem{G85} J.R. Gott, Ap. J. {\bf 288}, 422 (1985).

\bibitem{AFV} M. Aryal, L.H. Ford and A. Vilenkin, Phys. Rev. {\bf
D34}, 2263 (1986).

\bibitem{KS} N. Kaiser and A. Stebbins, Nature {\bf 310}, 391 (1984).

\bibitem{PS} D.D. Spergel and U. Pen, Phys. Rev. {\bf D51}, 4099
(1995).

\bibitem{Stebbins} A. Stebbins and S. Veeraraghavan, Phys. Rev. {\bf
D51}, 1465 (1995).

\bibitem{Refp2} See, e.g., Eq. (13.47) and Fig. 13.5 of Peebles' book
\cite{Peebles}.

\bibitem{Peebles} P.J.E. Peebles, {\it Principles of Physical
Cosmology} (Princeton University Press, Princeton, 1993). 

\bibitem{Ferreira} P. Ferreira, Phys. Rev. Lett. {\bf 74}, 3522
(1995).

\end{thebibliography}
\end{document}